\documentclass[a4paper,10pt,twocolumn]{article}
\usepackage{epsf,amsmath}
\usepackage{amssymb}
\usepackage{indentfirst}
\makeatletter
\def\@biblabel#1{#1.}
\makeatother

\begin{document}

\title{\Large \bf Modeling the Wind of the Be Star SS 2883}

\author{\bf A. I. Bogomazov\footnote{bogomazov@sai.msu.ru} \\
\it Moscow State University, Moscow, Russia \\}
\date{\begin{minipage}{15.5cm} \small
Observations of eclipses of the radio pulsar B1259-63 by the disk
of its Be-star companion SS 2883 provide an excellent opportunity
to study the winds of stars of this type. The eclipses lead to
variations in the radio flux (due to variations in the free-free
absorption), dispersion measure, rotation measure, and linear
polarization of the pulsar. We have carried out numerical modeling
of the parameters of the Be-star wind and compared the results
with observations. The analysis assumes that the Be-star wind has
two components: a disk wind in the equatorial plane of the Be star
with a power-law fall-off in the electron density $n_e$ with
distance from the center of the star $\rho$
($n_{e}\sim\rho^{-\beta_0}$), and a spherical wind above the
poles. The parameters for a disk model of the wind are estimated.
The disk is thin (opening angle $7.5^{\circ}$) and dense (electron
density at the stellar surface $n_{0e}\sim 10^{12}$/cm$^3$,
$\beta_0=2.55$), The spherical wind is weak ($n_{0e}\lesssim
10^{9}$/cm$^3$, $\beta_0=2$). \end{minipage} }

\maketitle \rm

\section{Introduction}

\indent The first binary system consisting of a radio pulsar and a
Be star was discovered more than ten years ago using the Parkes
radio telescope \cite{johnston1,johnston2}. This is the B1259-63
system, in which the pulsar moves along a very elongated orbit
($e\ge0.87$) around its companion -- the 10$^m$ Be star SS 2883.

The first theoretical estimates of the number of such systems,
carried out in 1983-1987 using the ``Scenario Machine''
\cite{korlip1}-\cite{lippro} showed that approximately
one in 700 observed radio pulsars should have an OB companion. The
detection of the B1259-63 system not only confirmed the possiblity
of the evolutionary scenario obtained by the authors of
\cite{korlip1}-\cite{lippro}, but also provided a
powerful tool for studies of the associated stellar wind, as was
also predicted in these studies. Currently, 1500 radio pulsars are
known, of which at least two are in binary systems with OB
companions (B1259-63 and J0045-7319), consistent with the above
estimate.

In 1997-1998, possible evolutionary tracks for the B1259-63 and
J0045-7319 systems were computed using the ``Scenario Machine''
\cite{raguzova}. These computations were based on an evolutionary
scenario that predicts the existence of systems consisting of a
radio pulsar and a massive optical component. The high
eccentricity of the orbit is explained as an effect of the
``kick'' given by the anisotropic supernova explosion that gave
rise to the pulsar. Possible magnitudes and the direction of the
kick velocity in the B1259-63 system were also derived
\cite{propos}.

In 2003, the ``Scenario Machine'' was used to estimate the number of
such systems in which the orbital plane of the pulsar and the
equatorial plane of the Be star do not coincide \cite{bogomazov}.
For characteristic anisotropic kick velocities of 50-200 km/s,
these two planes should be inclined relative to one another by
several tens of degrees in more than half of these systems.

A Be star is a main-sequence star of spectral class B which has
one or more Balmer emission lines in its spectrum, with these
lines usually displaying two peaks. Struve proposed in 1931 that
this spectral characteristic could be explained as radiation from
a rotating disk associated with the Be star. These disks have now
been observed in the optical, infrared, and radio. They are
comprised of dense, slowly rotating material that is located in or
near the equatorial plane of the Be star. In addition, Be stars
produce winds with low density and high velocity.

Observations of the transit of PSR B1259-63 provide a unique
opportunity for studying the characteristics of the disk of the Be
star based on the observed variations of the radio flux, linear
polarization, rotation measure, and pulse delay. This is possible
because the disk of SS 2883 is inclined relative to the orbital
plane of its companion, so that the pulsar is sometimes eclipsed
by the disk.

Attempts were undertaken to construct a model for the Be-star wind
and compare the calculated parameters with the observed dispersion
measure, rotation measure, and pulse delay time. A disk model with
an exponential fall-off of the electron density with distance from
the Be star and with height above the plane of the disk was
considered in \cite{melatos}.

A model with a power-law fall-off of the density in the disk is
more physically justified, however \cite{johnston3}; with some
parameters, this model was able to explain the observed variations
in the dispersion measure near periastron. A model with a clumpy
disk wind from the Be star has also been developed
\cite{johnston4,connors1} (in addition to, not replacing the
power-law disk). Rapidly moving bubbles (velocities of
$\approx2000$ km/s) with electron densities that differ from the
value in the surrounding region ($n_{e}\sim10^{6}/$cm$^{3}$, and
dimensions $<10^{10}$cm at a distance of 20-50 stellar radii) give
rise to appreciable fluctuations in the electron density along the
line of sight, and thereby to fluctuations in the flux, dispersion
measure, and rotation measure.

An attempt to establish the position of the orbital plane of PSR
B1259-63 relative to the equatorial plane of SS 2883 was made
using timing measurements for the pulsar \cite{wex,wang}. However,
it proved impossible to construct a unique model for the system due
to the noise level and the prolonged eclipses of the pulsating
radio emission.

A disk model for the Be-star wind is also considered, for example,
in \cite{araujo1,araujo2}, where the ratio $N$ of the densities of
the outflowing material in the equatorial plane and at the poles
is derived. If the Be star rotates with a velocity $v$ that is
90\% of the critical velocity $v_{crit}$ then $N=150$; slower
rotation lowers this value (when $v=70\%\cdot v_{crit}$, $N\approx
15-20$). According to\cite{porter}, the rotational velocity of SS
2883 is approximately 70\% of the critical velocity.

We consider here the absorption of the radio emission in a model
with a power-law disk wind in the equatorial plane and a spherical
wind above the poles of the Be star (without including the effect
of clumpiness). In contrast to previous studies, we compare the
calculated and observed fluxes of pulsating radio emission. The
derived parameters of the wind differ from those presented in
\cite{johnston3}. The available rotation measure data are
sufficient only to derive order of magnitude estimates of the
magnetic field strength in the wind (in particular, in bubbles),
as was done in \cite{johnston3}-\cite{connors1}. Accordingly, we
will consider further only the radio flux and dispersion measure.

\section{Model Be-star wind}

\renewcommand{\figurename}{Fig.}
\begin{figure}
\vspace{0cm}\hspace{0.5cm} \epsfxsize=0.4  \textwidth
\epsfbox{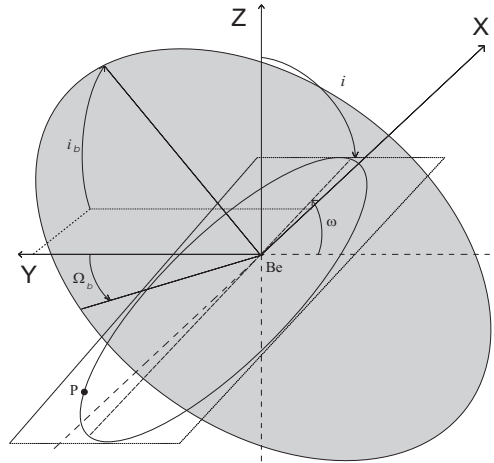} \vspace{0cm} \caption{ Schematic model for a
system consisting of a radio pulsar and a Be star. The X axis is
directed along the line of sight away form the observer. $\omega$
is the longitude of the ascending node, $i$ is the orbital
inclination (the angle between the plane of the sky YZ and the
orbital plane of the pulsar), P indicates the pulsar and Be the Be
star, $i_b$ is the angle between the XY plane and the plane of the
Be-star disk, and $\Omega_b$ is the angle between the Y axis and
the line corresponding to the intersection of the XY plane and the
disk of the Be star. } \label{risSys}
\end{figure}

\indent Figure \ref{risSys} depicts a schematic model of a binary
system consisting of a radio pulsar and a Be star. In this figure,
$\omega$ is the longitude of the ascending node, $i$ is the
orbital inclination (the angle between the plane of the sky YZ and
the orbital plane of the pulsar), P indicates the pulsar and Be
the Be star, i$_b$ is the angle between the XY plane and the plane
of the Be-star disk, and $\Omega_b$ is the angle between the Y
axis and the line corresponding to the intersection of the XY
plane and the disk of the Be star. The X axis is oriented along
the line of sight away from the observer.

\renewcommand{\figurename}{Fig.}
\begin{figure}
\vspace{0cm}\hspace{0cm} \epsfxsize=0.45 \textwidth
\epsfbox{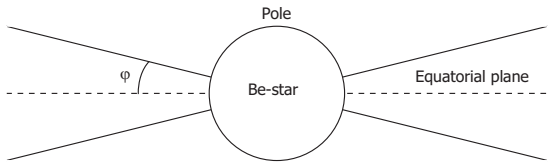} \vspace{0cm} \caption{ Model Be-star wind.
There is a disk wind with a power-law fall-off in the electron
density in the equatorial plane and a spherical wind above the
poles. } \label{risBe}
\end{figure}

The calculated model for the wind of SS 2883 is presented in Fig.
\ref{risBe}. There is a disk wind with opening angle $\varphi$ in
the equatorial plane, and a spherical wind above the poles, which
we take here to be isothermal and to have a constant radial speed.
We assume that the outflowing gas is fully ionized.

The electron density $n_e$ a distance $\rho$ from the star is
calculated using the formula

\begin{equation}
n_{e}=n_{0e}\left(\frac{\rho_{0}}{\rho}\right)^{\beta_{0}}.
\label{concentration}
\end{equation}

\noindent Here, $n_{0e}$ is the electron density at the stellar
surface (i.e., at a distance $\rho_0$ from its center). For the
spherical wind, $\beta_0=2$; this is a free parameter for the disk
wind, whose value is one of the parameters for which we are
searching.

We also assumed that the temperature of the disk wind can be
determined by the power-law relation

\begin{equation}
T=T_0\left(\frac{\rho_{0}}{\rho}\right)^{\beta_{1}}. \label{T}
\end{equation}

\noindent Here, $T$ is the temperature of the plasma at a distance $\rho$ from
the star and $T_0$ is the temperature at the stellar surface.

The optical depth to radio emission can be calculated using the
formula \cite{caplan}:

\begin{equation}
\tau=0.34\left(\frac{1GHz}{\nu}\right)^{2.1}\left(\frac{10^4K}{T}\right)^{1.35}\frac{EM}{10^6Pc\cdot
cm^{-6}}. \label{tau}
\end{equation}

\noindent Here, $\nu$ is the frequency of the radio emission, $T$
the temperature of the wind (\ref{T}), and EM the emission
measure, which is defined to be

\begin{equation}
EM=\int n_e^2 dl \label{EM}.
\end{equation}

The dispersion measure is given by the formula

\begin{equation}
DM=\int n_edl \label{DM}.
\end{equation}

\noindent The quantity $n_e$ in (\ref{EM}) and (\ref{DM}) is
calculated using (\ref{concentration}).

\renewcommand{\figurename}{Fig.}
\begin{figure}
\vspace{0cm}\hspace{0cm} \epsfxsize=0.48 \textwidth
\epsfbox{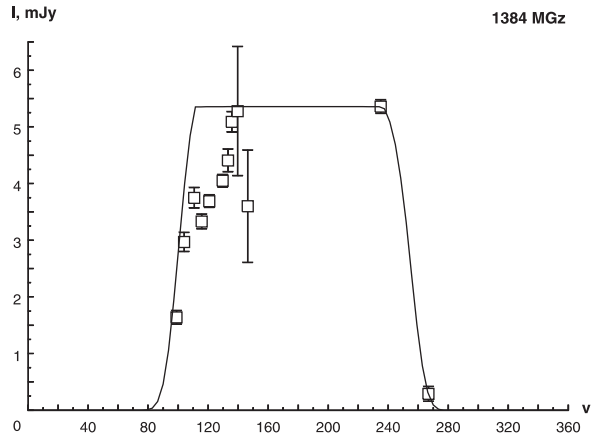} \vspace{0cm} \caption{ Dependence of the
1384 MHz radio flux of B1259-63 on the position of the pulsar
relative to SS 2883; $v$ is the true anomaly. The hollow squares
show the observational data with their errors. The solid curve
denotes the calculated model. } \label{ris1384}
\end{figure}

\section{Results}

\indent We carried out theoretical calculations of the flux and
dispersion measure of the radio emission of the B1259-63 pulsar as
a function of its position relative to its companion -- the Be
star SS 2883. These calculated values were compared with the
observations.

The observational data on the radio fluxes were taken from Table 2
of \cite{connors1}, and on the dispersion measures from Table 1 of
\cite{melatos}, Table 2 of \cite{johnston3}, and Table 3 of
\cite{connors1}. The longitude of the ascending node of the pulsar
orbit in the B1259-63 system is $\omega=138^{\circ}$
\cite{johnston2}, the orbital inclination is $i=36^{\circ}$, the
characteristic radius of the star is $\rho_0=6R_{\odot}$ the mass
of the optical star is $M=10M_{\odot}$ and the projection of the
semi-major axis onto the line of sight is $a \sin$i$=1295.98$
light seconds \cite{johnston2}. The interstellar contribution to
the dispersion measure of the B1259-63 pulsar is 146.8~Pc/cm$^3$
\cite{connors1}.

\renewcommand{\figurename}{Fig.}
\begin{figure}
\vspace{0cm}\hspace{0cm} \epsfxsize=0.48 \textwidth
\epsfbox{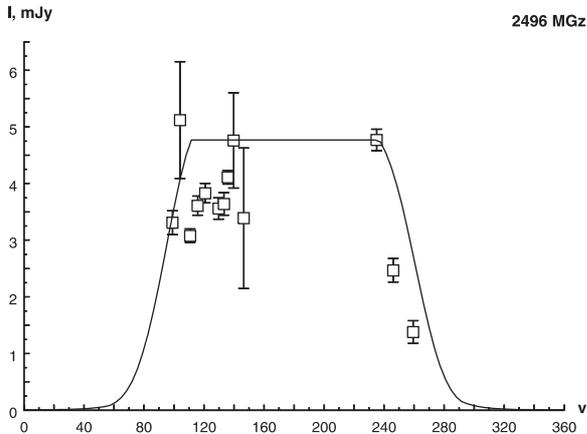} \vspace{0cm} \caption{ Same as Fig. 3 for
2496 MHz. } \label{ris2496}
\end{figure}

The velocity of the Be-star wind at infinity is $V_\infty$ ðàâíà
$1000-1300$ km/s in the equatorial plane and, $\approx 2300$ km/s
at the poles \cite{araujo2} (the characteristic wind velocity at
infinity for SS 2883 is $\approx 1350 \pm 200$ km/s
\cite{collum}). The radial outflow velocity of the disk material
at the Be-star surface is 5-10 km/s \cite{waters}.

To check the resulting parameters of the stellar wind, we must
compare the mass-loss rate in the model obtained with the
characteristic values for Be stars. The rate at which matter flows
from the star can be found from the expression

\begin{equation}
\dot M =\Omega \rho^2 n m_i V; \label{Mloss}
\end{equation}

\noindent where $\rho$ is the distance from the star, $n$ the
particle density of the wind at the distance $\rho$ which is
calculated using (\ref{concentration}), $m_i$ the mean mass of the
ions in the wind, $V$ the wind velocity at the distance $\rho$,
and $\Omega$ the solid angle into which the given type of wind
(spherical or disk) flows.

\renewcommand{\figurename}{Fig.}
\begin{figure}
\vspace{0cm}\hspace{0cm} \epsfxsize=0.48 \textwidth
\epsfbox{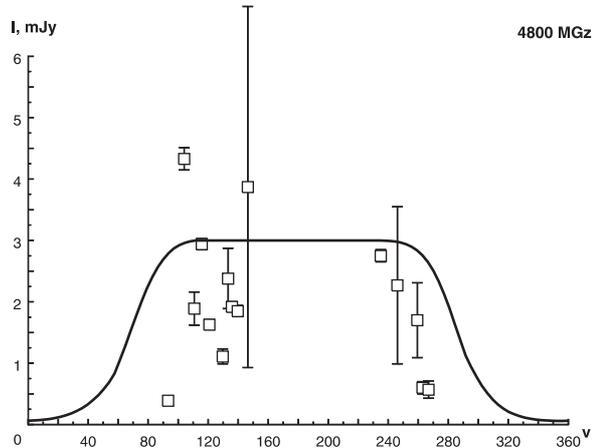} \vspace{0cm} \caption{ Same as Fig. 3 for
4800 MHz. } \label{ris4800}
\end{figure}

The mass loss of Be stars was investigated in detail in
\cite{tutukov}. A Be star with an initial mass of $10 M_{\odot}$
loses mass at the rate $\dot M=3\cdot 10^{-9}M_{\odot}$/year.

We carried out numerical computations in order to determine the
best-fit orientation of the disk of the Be star SS 2883 relative
to the orbital plane of PSR B1259-63, which is given by the
parameters $\Omega_b$ and $i_b$ (Fig. \ref{risSys}). We also
wished to determine the best-fit values of $\beta_0$ for the disk
wind and the electron density at the stellar surface $n_0$ for
both the disk and spherical winds ($\beta_0=2$ for the spherical
wind). We constructed theoretical light curves for the pulsar at
1384, 2496, 4800, and 8400 MHz together with the dependence of the
dispersion measure of the pulsar radio emission on the position of
the pulsar relative to its companion.

\begin{figure}
\vspace{0cm}\hspace{0cm} \epsfxsize=0.48 \textwidth
\epsfbox{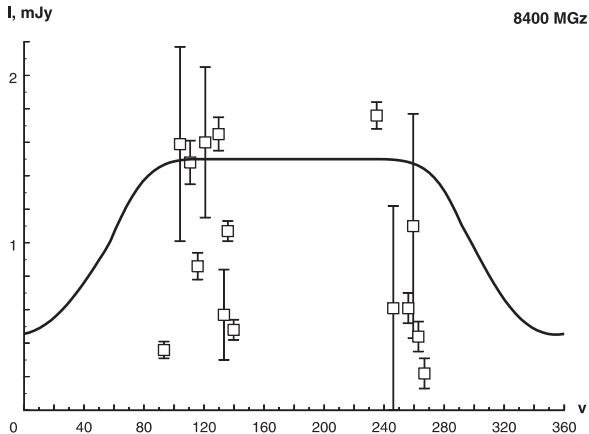} \vspace{0cm} \caption{ Same as Fig. 3 for
8400 MHz. } \label{ris8400}
\end{figure}

\begin{figure}
\vspace{0cm}\hspace{0cm} \epsfxsize=0.48 \textwidth
\epsfbox{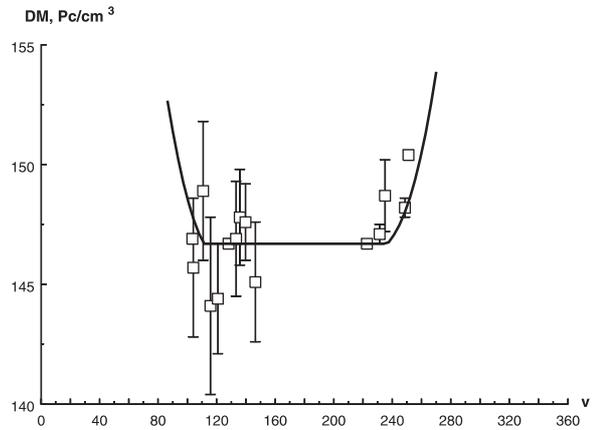} \vspace{0cm} \caption{ Dependence of the
dispersion measure on the true anomaly $v$ of the B1259-63 pulsar.
The hollow squares indicate the observational data with their
errors. The solid curve denotes the model with a thin and dense
disk wind and a negligibly small spherical wind (we took the
interstellar contribution to the dispersion measure to be 146.8
pc/cm3). } \label{risDM}
\end{figure}

The results of the numerical simulations are shown in
\ref{ris1384} - \ref{risDM} together with the observational data.
The criterion used to evaluate the correctness of a model is the
closeness of the computed curves to the observational data. In
addition, the model should not contradict our understanding of the
matter outflow rate in Be stars.

\renewcommand{\figurename}{Fig.}

The computations yielded the following best-fit parameters for the
disk model. The disk opening angle was $\varphi=7.5^{\circ}$, the
electron density in the disk at the stellar surface was
$n_{0e}\approx 1\cdot 10^{12}$/cm$^3$, and $\beta_0=2.55$. The
spherical wind was weak ($n_{0e}\lesssim 10^{9}$/cm$^3$,
$\beta_0=2$). The orientation of the SS 2883 disk relative to the
orbital plane is described by the parameters $\Omega_b$ and i$_b$
(Fig. \ref{risSys}), which have the values $\Omega_b=12^{\circ}$
and $i_b=67^{\circ}$ in the best-fit model. The disappearance of
the pulsating emission of B1259-63 near periastron is due to
eclipses of the pulsar by the Be-star disk. The weak spherical
wind does not make an appreciable contribution to either the
decrease in the radio flux or the increase in the dispersion
measure. The model mass-loss rate via the SS 2883 wind estimated
using (\ref{Mloss}) is consistent with our current understanding
of the mass-loss rates of Be stars: $\dot M\approx 3\cdot
10^{-9}M_{\odot}$/yr, in good agreement with the results of
\cite{tutukov}.

Although we have assumed that the temperature of the wind may
depend on the distance from the Be star (\ref{T}), the best-fit
model was obtained for a constant temperature, equal to $10^4$K.
Substituting (\ref{concentration}) and (\ref{T}) into (\ref{tau})
we can see that the behavior of the light curve will not change if
$2\beta_0-1.35\beta_1=5.1$ if we suppose that $\beta_1\ne 0$ (so
that consequently, $\beta_0\ne 2.55$). However, the variations of
$\beta_0$ must not be in contradiction with the observed
dispersion measures, and cannot be large.

Figure \ref{ris1384} clearly displays a secondary minimum of the
light curve at values of the true anomaly $v$ from $100^{\circ}$
to $120^{\circ}$. Our model for the wind is not able to fit this
observational feature. The computations yielded a model with
$\Omega_b\approx -55^{\circ}$ that can fit this secondary minimum,
whose presence is then explained by the absorption of the radio
emission in the disk. However, in this case, the pulsar and
observer are located on the same side of the disk at periastron,
and we must introduce a strong spherical wind in order to fit the
decrease in the radio flux. This led to an implausibly high
mass-loss rate ($\dot M\approx 3\cdot 10^{-7}M_{\odot}$/yr). In
addition, a strong spherical wind would make an appreciable
contribution to the dispersion measure, so that it would be
necessary to appreciably lower the interstellar dispersion measure
(from 146.8 PC/cm$^3$ to 135-140 Pc/cm$^3$). We, accordingly,
rejected this second model.

The parameters of the best-fit model presented here differ
substantially from the results of \cite{johnston3}, where, in
particular, it was found that $\beta_0=4.2$ in the disk and that
the mass-loss rate of the optical star is $\dot M=5\cdot
10^{-8}M_{\odot}$/yr.

\section{Conclusion}

\renewcommand{\figurename}{Fig.}

\indent Our computations have yielded the following best fit
paramters for the disk model used. The opening angle of the disk
is $\varphi=7.5^{\circ}$, the electron density in the disk at the
stellar surface is $n_{0e}\approx 1\cdot 10^{12}$/cm$^3$, ïàðàìåòð
$\beta_0=2.55$. The spherical wind is weak ($n_{0e}\lesssim
10^{9}$/cm$^3$, $\beta_0=2$) and cannot make a significant
contribution to either the decrease in the radio flux or the
increase in the dispersion measure near periastron. The
orientation of the SS 2883 disk relative to the pulsar's orbital
plane is fit by the parameters $\Omega_b=12^{\circ}$ and
$i_b=67^{\circ}$ (Fig. \ref{risSys}). The mass-loss rate in the SS
2883 wind estimated using (\ref{Mloss}), is consistent with
current thinking about the mass-loss rates of Be stars: $\dot
M\approx 3\cdot 10^{-9}M_{\odot}$/yr, in good agreement with the
results of \cite{tutukov}. The disappearance of the pulsating
radio emission of B1259-63 near periastron is due to eclipsing of
the pulsar by the dense disk of the Be star.

The author thanks the anonymous referee for useful comments.

\end{document}